\documentclass[twocolumn]{svjour3}

\usepackage{amsmath}
\usepackage{amssymb}
\usepackage{array}
\usepackage{booktabs}
\usepackage{graphicx}
\usepackage{url}

\graphicspath{{./}}

\newcommand{\ii}{\mathrm{i}}

\journalname{Eur. Phys. J. B}

\begin{document}

\title{Van Hemmen interactions in a one-dimensional swarmalator model}

\author{Kevin O'Keeffe}

\institute{Kevin O'Keeffe \at
           Starling Research Institute \\
           \email{kevin.p.okeeffe@gmail.com}}

\date{Received: date / Accepted: date}

\maketitle

\begin{abstract}
We study a one-dimensional swarmalator model with van Hemmen pair disorder in the phase coupling.
Pair disorder has two effects.
First, it splits the static ring-model states into sync, split, splay, and phase-wave branches organized by the rainbow order parameters $r,s$ and four sign-weighted glass order parameters.
Second, because the oscillators move, it creates active macrostates absent from the immobile Kuramoto-van Hemmen model: a bursty active async state and a glassy phase wave with rotating glass order.
For balanced sign patterns we derive a six-field reduction, the exact finite-$N$ sync boundary, a closed first split branch with its first spatial destabilization, and an exact antiphase phase-wave branch.
The iid sign audits preserve the tested state ordering but shift finite-$N$ thresholds through sample imbalance.
The remaining challenge is a nonlinear theory of the two active branches.
\end{abstract}

\section{Introduction}

Swarmalators are mobile oscillators whose spatial and phase dynamics are coupled.
They model systems in which synchronization and self-assembly feed back on one another, from active colloids and biological microswimmers to magnetic domain walls \cite{o2017oscillators,quillen2021metachronal,yan2012linking,hrabec2018velocity}.

The one-dimensional ring model has become the analytically cleanest setting for this feedback.
Its async, phase-wave, and sync states have exact finite-$N$ and continuum stability results \cite{o2022collective,yoon2022sync,global_sync,o2025stability}.
The same framework now supports distributed coupling strengths, attractive and repulsive interactions, pinning, thermal noise, forcing, finite interaction range, delay, self-propulsion, inertia, pulsatile coupling, and higher-order interactions \cite{o2022swarmalators,hao2023attractive,sar2023pinning,sar2023swarmalators,sar2024solvable,hong2023swarmalators,anwar2024forced,sar2025effects,blum2024swarmalators,okeeffe2026time_delay,okeeffe2026selfprop,okeeffe2026inertia,ghosh2025dynamics,anwar2024collective}.
Complementary work has explored coupling disorder in related swarmalator models \cite{hong2021coupling}.

\begin{figure}[!htbp]
\centering
\includegraphics[width=0.75\columnwidth]{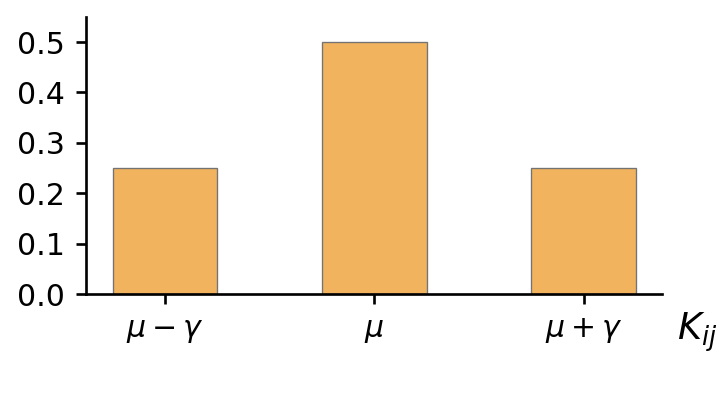}
\caption{Van Hemmen coupling distribution in Eq.~\eqref{eq:vh}. For balanced random signs, the interaction values concentrate at $\mu-\gamma$, $\mu$, and $\mu+\gamma$ with weights $1/4$, $1/2$, and $1/4$.}
\label{fig:coupling}
\end{figure}

What is still missing is a solvable ring theory for pair disorder.
Site disorder attaches a random parameter to each swarmalator; pair disorder attaches it to an interaction.
This is the natural language for contact, hydrodynamic, magnetic, or chemical couplings whose strength depends on both units, and it raises a question that site disorder does not: can quenched randomness in the links create glass order while the oscillators continue to move?
We use the analytically structured van Hemmen form
\begin{align}
    K_{ij}=\mu+\frac{\gamma}{2}(p_iq_j+p_jq_i),
    \label{eq:vh}
\end{align}
with quenched signs $p_i,q_i=\pm1$.
For balanced signs, one quarter of the interactions are $\mu-\gamma$, one half are $\mu$, and one quarter are $\mu+\gamma$; Fig.~\ref{fig:coupling} shows the resulting three-point distribution.
Van Hemmen introduced this low-rank structure as a solvable spin glass \cite{vanHemmen1982}.
With a finite Lorentzian frequency width, the corresponding Kuramoto model produces incoherent, partially synchronized, antiphase, and mixed states \cite{kloumann2013phase}.
In the swarmalator model the same disorder also drives motion: phases reshape future spatial interactions, and positions reshape future phase interactions.

Our main result is that pair disorder creates both static glass order and active macroscopic motion.
For clean balanced signs we find a static sequence sync $\to$ split $\to$ splay $\to$ phase wave, plus two active regimes: bursty async and a glassy phase wave.
For iid signs, finite-size imbalance shifts thresholds and adds a small glass background to sync, but the tested state ordering persists.

\section{Model}

We set the spatial coupling to one and put the van Hemmen disorder in the phase coupling:
\begin{align}
    \dot{x}_i &=
    \frac{1}{N}\sum_{j=1}^N \sin(x_j-x_i)\cos(\theta_j-\theta_i),
    \label{eq:model-x}\\
    \dot{\theta}_i &=
    \frac{1}{N}\sum_{j=1}^N K_{ij}\sin(\theta_j-\theta_i)\cos(x_j-x_i),
    \label{eq:model-theta}
\end{align}
where $x_i,\theta_i\in S^1$ and $K_{ij}$ is given by Eq.~\eqref{eq:vh}.
Natural speeds and frequencies are set to zero so that the only source of frustration is the coupling disorder.

The useful coordinates are
\begin{align}
    \xi_i=x_i+\theta_i,\qquad \eta_i=x_i-\theta_i.
\end{align}
We monitor the rainbow order parameters
\begin{align}
    r e^{\ii\Phi_+}
        &=\frac{1}{N}\sum_{j=1}^N e^{\ii\xi_j},\\
    s e^{\ii\Phi_-}
        &=\frac{1}{N}\sum_{j=1}^N e^{\ii\eta_j},
\end{align}
and the four glass order parameters
\begin{align}
    u_\pm e^{\ii\Psi_\pm}
        &=\frac{1}{N}\sum_{j=1}^N p_j e^{\ii(x_j\pm\theta_j)},\\
    v_\pm e^{\ii\Theta_\pm}
        &=\frac{1}{N}\sum_{j=1}^N q_j e^{\ii(x_j\pm\theta_j)}.
\end{align}
The rainbow order parameters are those of the ordinary ring model.
The glass order parameters measure whether the spatial-phase pattern is correlated with the quenched van Hemmen signs.

Let
\begin{align}
    F_i^+ &= \operatorname{Im}(r e^{\ii\Phi_+}e^{-\ii\xi_i}),\\
    F_i^- &= \operatorname{Im}(s e^{\ii\Phi_-}e^{-\ii\eta_i}),\\
    G_i^+ &= \operatorname{Im}\{(p_i v_+e^{\ii\Theta_+}
        +q_i u_+e^{\ii\Psi_+})e^{-\ii\xi_i}\},\\
    G_i^- &= \operatorname{Im}\{(p_i v_-e^{\ii\Theta_-}
        +q_i u_-e^{\ii\Psi_-})e^{-\ii\eta_i}\}.
\end{align}
Then Eqs.~\eqref{eq:model-x}--\eqref{eq:model-theta} become
\begin{align}
    \dot{\xi}_i &=
        \frac{1+\mu}{2}F_i^+
        +\frac{1-\mu}{2}F_i^-
        +\frac{\gamma}{4}(G_i^+-G_i^-),
        \label{eq:xi}\\
    \dot{\eta}_i &=
        \frac{1-\mu}{2}F_i^+
        +\frac{1+\mu}{2}F_i^-
        -\frac{\gamma}{4}G_i^+
        +\frac{\gamma}{4}G_i^-.
        \label{eq:eta}
\end{align}
Thus the disorder enters through only four additional mean fields.
When $\gamma=0$, the model reduces to the ordinary one-dimensional ring model with transformed self-coupling $(1+\mu)/2$ and cross-coupling $(1-\mu)/2$.

\section{Numerical Results}

We integrated Eqs.~\eqref{eq:model-x}--\eqref{eq:model-theta} with a fourth-order Runge-Kutta method from random initial conditions and used the speed diagnostic $m(t)=N^{-1}\sum_{i=1}^N(\dot x_i^2+\dot\theta_i^2)^{1/2}$.

\subsection{Balanced Disorder}

\begin{figure*}[!t]
\centering
\includegraphics[width=0.95\textwidth]{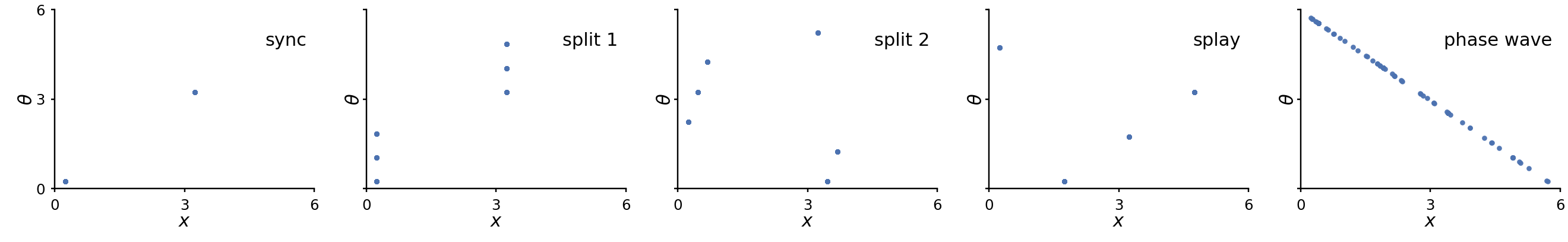}
\caption{Static states of the van Hemmen swarmalator model in the $(x,\theta)$ plane. From left to right the states are sync, split 1, split 2, splay, and phase wave. The sync and split-1 panels use the exact states; the numerical panels use $(\mu,\gamma)=(1,3.2)$, $(1,5)$, and $(-0.75,4)$, respectively, with $N=100$ and $dt=0.1$.}
\label{fig:static-states}
\end{figure*}

Unless stated otherwise, the figures use the prepared balanced sign pattern
$p_i=(-1)^i$ and $q_i=(-1)^{i(i-1)/2}$.
For $N$ divisible by four, the four $(p_i,q_i)$ groups have equal populations and overlap $\chi=0$.
This prepared pattern is the clean balanced problem: it preserves the three van Hemmen interaction values and removes the $O(N^{-1/2})$ sample imbalance of a fully random draw.
Section~\ref{sec:iid-robustness} then checks what changes for iid signs.
Figure~\ref{fig:static-states} shows representative balanced-disorder attractors in the $(x,\theta)$ plane.

The static states follow a simple progression.
Sync has $r=s=1$ and no glass order.
Split states lower the rainbow order and turn on glass order; in split 2 the $+$ and $-$ sectors no longer coincide.
Splay keeps one transformed coordinate clustered and spreads the other across sign groups.
The phase wave spreads one transformed coordinate around the circle.

\begin{figure}[tbp]
\centering
\includegraphics[width=0.92\columnwidth]{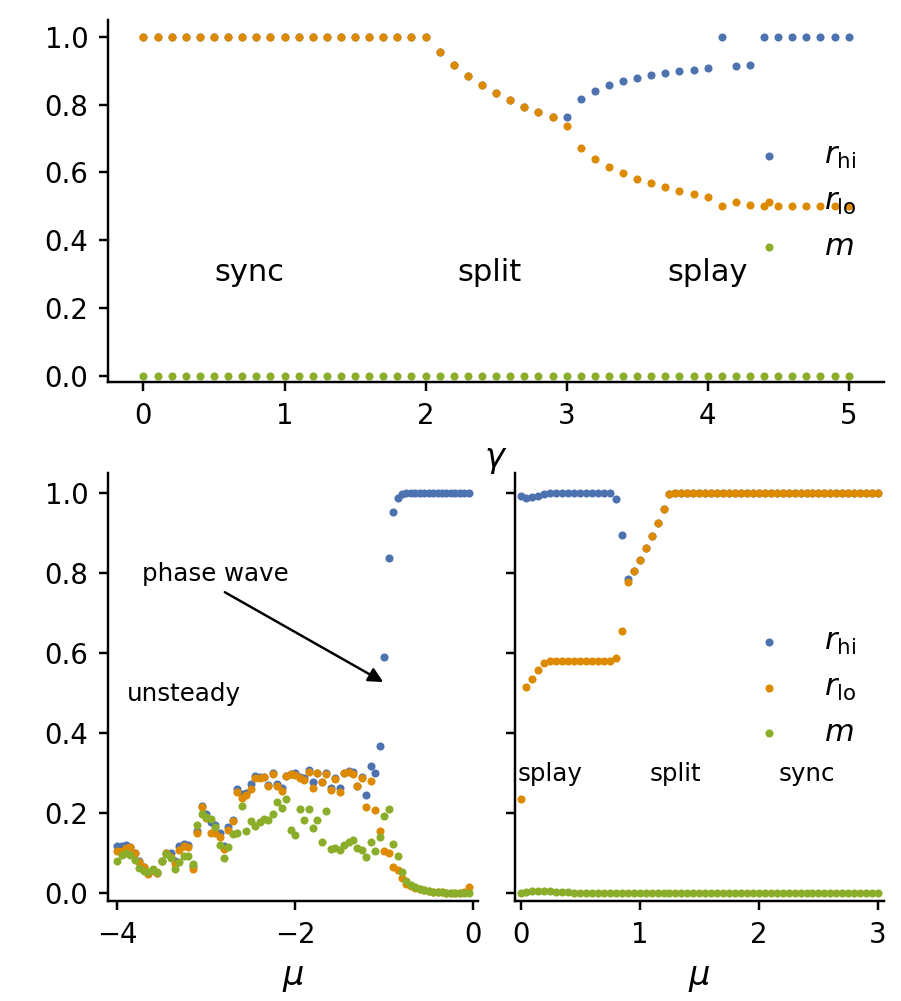}
\caption{Order parameters across the observed transitions. Top: $\mu=1$ and $\gamma$ is varied. Bottom: $\gamma=4$ and $\mu$ is varied, with a zoom of the positive-$\mu$ static branches. The blue and orange points are $r_{\rm hi}=\max(r,s)$ and $r_{\rm lo}=\min(r,s)$; the green points show the mean speed $m$. In the unsteady regime, these quantities are averaged over the final part of the simulation. Simulation details: $(dt,T,N)=(0.1,300,100)$.}
\label{fig:order-pars}
\end{figure}

Figure~\ref{fig:order-pars} tracks the sequence with $r_{\rm hi}=\max(r,s)$, $r_{\rm lo}=\min(r,s)$, and the mean speed $m$.
At fixed positive $\mu$, increasing $\gamma$ drives sync $\to$ split $\to$ splay.
At fixed large $\gamma$, sweeping $\mu$ gives active async, phase wave, splay, split, and sync.
Nonzero $m$ marks the active states.

Active async is the clearest sign that mobility changes the Kuramoto-van Hemmen story.
For $\gamma=0$ and sufficiently negative $\mu$, the ordinary ring model is static async.
With van Hemmen disorder, the rainbow order remains small on average but bursts irregularly away from zero; see Fig.~\ref{fig:bursty}.

Near the phase-wave branch there is a more organized unsteady state, the glassy phase wave.
Let ``hi'' denote the transformed coordinate with larger rainbow order and ``lo'' the complementary coordinate.
As shown in Fig.~\ref{fig:glassy}, $r_{\rm hi}$ grows slowly toward a large value, $r_{\rm lo}$ stays small, and the glass order parameters $u_{\rm lo}$ and $v_{\rm lo}$ quickly approach a nonzero value.
Their phases rotate while their magnitudes remain nearly steady.
This time-scale separation is absent from the ordinary identical-frequency ring model.

Appendix~\ref{app:validation} shows the same signature in prepared-sign sweeps.
The state is phase-wave-like in the transformed coordinates, but the spread coordinate carries sign-weighted order.
Its mean speed is nonzero because the glass phases rotate.

\subsection{Robustness to iid Signs}
\label{sec:iid-robustness}

The prepared signs isolate the balanced van Hemmen structure from finite-size sample imbalance.
For iid signs, imbalance enters through the overlap $\chi=N^{-1}p^Tq$ and through nonzero means $\langle p\rangle,\langle q\rangle$.
The first shifts the sync threshold; the second gives sync an $O(N^{-1/2})$ glass background.

We ran two iid checks.
An audit of twenty iid sign draws gave finite-$N$ linear sync thresholds in the ranges $1.73<\gamma_c<3.56$ for $N=64$, $1.69<\gamma_c<2.50$ for $N=100$, and $1.71<\gamma_c<2.46$ for $N=128$ at $\mu=1$.
A separate random-start sweep for three iid sign draws at $N=64$ gave the tail-averaged ranges in Table~\ref{tab:iid-ranges}, where $G_{\rm hi}=\max\{(u_++v_+)/2,(u_-+v_-)/2\}$.

\begin{table}[tbp]
\caption{Tail-averaged ranges from three iid sign realizations at $\mu=1$ and $N=64$.}
\label{tab:iid-ranges}
\centering
\footnotesize
\begin{tabular}{ccccc}
\hline\noalign{\smallskip}
$\gamma$ & $m$ & $r_{\rm hi}$ & $r_{\rm lo}$ & $G_{\rm hi}$ \\
\noalign{\smallskip}\hline\noalign{\smallskip}
$0$ & $<4\times10^{-15}$ & $1.00$ & $1.00$ & $0.016$--$0.156$ \\
$1$ & $<4\times10^{-15}$ & $1.00$ & $1.00$ & $0.016$--$0.156$ \\
$2.5$ & $<3\times10^{-15}$ & $0.807$--$0.938$ & $0.807$--$0.938$ & $0.275$--$0.403$ \\
$4.5$ & $<1.1\times10^{-7}$ & $1.00$ & $0.500$--$0.656$ & $0.438$--$0.516$ \\
\noalign{\smallskip}\hline
\end{tabular}
\end{table}

The iid draws gave sync-like order at $\gamma=0,1$, split-like glass order at $\gamma=2.5$, and a splay-like state at $\gamma=4.5$.
Thus the tested iid realizations preserve the balanced state sequence, but not the finite-$N$ transition locations.

\section{Analysis}

\subsection{Sync}

\begin{figure}[!htbp]
\centering
\includegraphics[width=0.8\columnwidth]{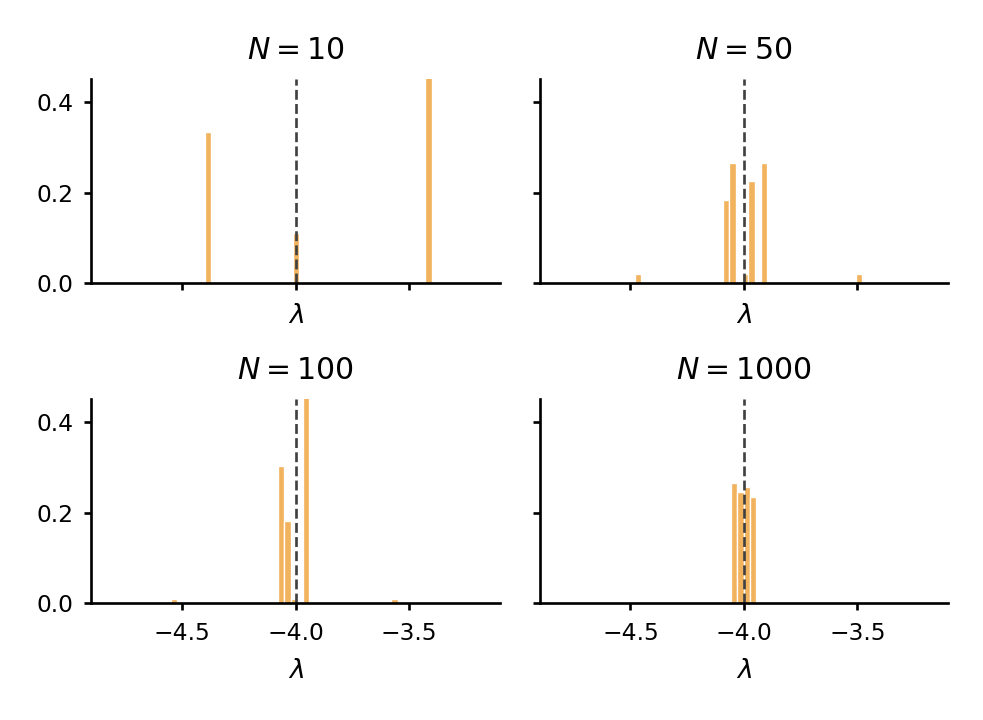}
\caption{Finite-size scaling of the phase-block spectrum at sync. The histograms show the nonneutral eigenvalues for one iid sign realization at $(\mu,\gamma)=(4,1)$; the dashed line is $-\mu$. The bulk narrows around $-\mu$ as $N$ grows, while the low-rank van Hemmen part controls the first instability.}
\label{fig:lam-sync}
\end{figure}

The sync state has $x_i=C_x$ and $\theta_i=C_\theta$ up to the usual $\pi$ symmetry.
Linearizing Eqs.~\eqref{eq:model-x}--\eqref{eq:model-theta} about one sync dot gives a block diagonal Jacobian,
\begin{align}
    M=\begin{pmatrix}A&0\\0&B\end{pmatrix}.
\end{align}
The spatial block is the complete-graph Laplacian
\begin{align}
    A_{ij}=
    \begin{cases}
    -(N-1)/N, & i=j,\\
    1/N, & i\ne j,
    \end{cases}
\end{align}
with eigenvalues $0$ and $-1$ with multiplicities $1$ and $N-1$.
The phase block is the weighted Laplacian
\begin{align}
    B_{ij}=
    \begin{cases}
    -N^{-1}\sum_{k\ne i}K_{ik}, & i=j,\\
    N^{-1}K_{ij}, & i\ne j.
    \end{cases}
    \label{eq:B}
\end{align}
It always has a neutral eigenvalue from phase rotation.
The remaining spectrum can be read off from the low-rank form of the van Hemmen matrix.
Let $\mathcal{K}$ denote the full matrix with entries $K_{ij}$, including the diagonal.
Then Eq.~\eqref{eq:B} can be written as
\begin{align}
    B=\frac{1}{N}\left(\mathcal{K}-\operatorname{diag}(\mathcal{K}{\bf 1})\right).
\end{align}
For an exactly balanced realization, $\sum_i p_i=\sum_iq_i=0$, so $\mathcal{K}{\bf 1}=N\mu{\bf 1}$ and
\begin{align}
    B=-\mu I+\frac{\mu}{N}{\bf 1}{\bf 1}^{T}
      +\frac{\gamma}{2N}\left(pq^T+qp^T\right).
    \label{eq:sync-low-rank}
\end{align}
Thus all vectors orthogonal to ${\bf 1}$, $p$, and $q$ have eigenvalue $-\mu$.
The two remaining nonneutral eigenvalues lie in the span of $p$ and $q$:
\begin{align}
    \lambda_+ &= -\mu+\frac{\gamma}{2}(1+\chi),\\
    \lambda_- &= -\mu-\frac{\gamma}{2}(1-\chi),
    \label{eq:sync-outliers}
\end{align}
where $\chi=N^{-1}p^Tq$ is the overlap of the two sign vectors.
For positive disorder strength, the first unstable eigenvalue is $\lambda_+$.
The finite-$N$ sync boundary for a balanced realization is therefore
\begin{align}
    \gamma_c(\chi)=\frac{2\mu}{1+\chi},\qquad \mu>0.
    \label{eq:sync-boundary}
\end{align}
For independent balanced signs, $\chi=O(N^{-1/2})$, so $\gamma_c\to2\mu$ as $N\to\infty$.
This is the first boundary seen in the top panel of Fig.~\ref{fig:order-pars}: at fixed $\mu=1$, the sync state loses stability near $\gamma=2$ and gives way to a split state.
Figure~\ref{fig:lam-sync} shows the finite-size collapse of the bulk spectrum.

\subsection{Split and Splay States}

The split and splay states are born from the sync instability in Eq.~\eqref{eq:sync-boundary}.
They are fixed points of the full $2N$-dimensional system, but their structure is low-dimensional because the signs $p_i,q_i$ divide the population into four quenched subgroups.
For a balanced realization, those subgroups are
\begin{align}
    (p_i,q_i)=(+,+),\ (+,-),\ (-,+),\ (-,-).
\end{align}
Let $a$ index these groups, with signs $(p_a,q_a)$, fraction $\rho_a$, and group coordinates $(X_a,\Theta_a)$.
The manifold on which all members of a group coincide is invariant and obeys
\begin{align}
    \dot X_a &=
    \sum_b \rho_b
    \sin(X_b-X_a)\cos(\Theta_b-\Theta_a),
    \label{eq:four-group-x}\\
    \dot\Theta_a &=
    \sum_b \rho_b
    \left[\mu+\frac{\gamma}{2}(p_aq_b+p_bq_a)\right]\notag\\
    &\quad{}\times
    \sin(\Theta_b-\Theta_a)\cos(X_b-X_a).
    \label{eq:four-group-theta}
\end{align}
For prepared balanced signs, $\rho_a=1/4$.
In transformed coordinates the same quotient is cleaner.
Let $\zeta_a^\pm$ denote $\xi_a$ and $\eta_a$.
The group order parameters are
\begin{align}
    r e^{\ii\Phi_+}&=\sum_b\rho_b e^{\ii\zeta_b^+},\qquad
    s e^{\ii\Phi_-}=\sum_b\rho_b e^{\ii\zeta_b^-},\notag\\
    u_\pm e^{\ii\Psi_\pm}
        &=\sum_b\rho_b p_b e^{\ii\zeta_b^\pm},\qquad
    v_\pm e^{\ii\Theta_\pm}
        =\sum_b\rho_b q_b e^{\ii\zeta_b^\pm}.
    \label{eq:four-group-order}
\end{align}
Then Eqs.~\eqref{eq:xi}--\eqref{eq:eta} hold with the particle index $i$ replaced by the group index $a$.
Thus the later split and splay branches are not arbitrary $2N$-dimensional fixed points; for prepared balanced signs they are roots of an eight-dimensional algebraic problem, minus the two neutral shifts.

The first split branch has an especially simple symmetric form in the transformed coordinates.
Let $s_a=(p_a+q_a)/2$, so $s_a=1$ for the $(+,+)$ group, $s_a=0$ for the two mixed groups, and $s_a=-1$ for the $(-,-)$ group.
Up to the two neutral shifts,
\begin{align}
    \xi_a=\Xi-s_a\delta,\qquad
    \eta_a=H+s_a\delta .
    \label{eq:split-form}
\end{align}
The two mixed sign groups are equivalent, so the four sign groups produce three transformed-coordinate levels rather than four.
In the original $(x,\theta)$ variables each level appears as a $\pi$-paired copy because $(x,\theta)$ and $(x+\pi,\theta+\pi)$ have the same $(\xi,\eta)$.
On this ansatz,
\begin{align}
    r e^{\ii\Phi_+}
        &=e^{\ii\Xi}\frac{1+\cos\delta}{2},\notag\\
    s e^{\ii\Phi_-}
        &=e^{\ii H}\frac{1+\cos\delta}{2},\notag\\
    u_+ e^{\ii\Psi_+}
        &=v_+ e^{\ii\Theta_+}
        =-\frac{\ii}{2}e^{\ii\Xi}\sin\delta,\notag\\
    u_- e^{\ii\Psi_-}
        &=v_- e^{\ii\Theta_-}
        =\frac{\ii}{2}e^{\ii H}\sin\delta .
    \label{eq:split-fields}
\end{align}
The mixed sign groups have $s_a=0$ and therefore zero residual automatically.
For the $s_a=\pm1$ groups, Eqs.~\eqref{eq:xi}--\eqref{eq:eta} reduce to the same scalar condition,
\begin{align}
    s_a\sin\delta
    \left[
    \frac{\mu}{2}(1+\cos\delta)
    -\frac{\gamma}{2}\cos\delta
    \right]=0 .
\end{align}
The nontrivial branch is therefore
\begin{align}
    \cos\delta=\frac{\mu}{\gamma-\mu}.
    \label{eq:split-branch}
\end{align}
For $\mu>0$, this branch exists for $\gamma\ge2\mu$, so it is born at the sync boundary in Eq.~\eqref{eq:sync-boundary} when $\chi=0$.
Along this branch,
\begin{align}
    r=s=\frac{1+\cos\delta}{2},
    \qquad
    u_\pm=v_\pm=\frac{\lvert\sin\delta\rvert}{2}.
    \label{eq:split-order}
\end{align}
For the representative split point in Fig.~\ref{fig:static-states} at $(\mu,\gamma)=(1,2.5)$, direct substitution into the full $2N$ equations gives a maximum residual speed below $10^{-14}$.
Equation~\eqref{eq:split-branch} gives $\cos\delta=2/3$, hence $r=s=5/6$ and $u_\pm=v_\pm=\sqrt{5}/6$, matching the stored state.
The full finite-$N$ Jacobian at the same point has only the two neutral modes associated with global shifts of $x$ and $\theta$; the largest nonneutral real part is $-0.278$.
This is a stable numerical fixed point, not a slowly moving transient.

The same branch also gives an exact prediction for where split 1 first becomes unstable.
In this solution all sign groups have the same position, $x_a=\mathrm{const.}$, while $\theta_a=-s_a\delta$.
Perturb only the positions by $\delta x_a\propto s_a$ and keep $\delta\theta_a=0$.
The spatial linearization gives
\begin{align}
    \lambda_{\rm sp}
    =\frac{1+\cos\delta}{2}\left(1-2\cos\delta\right).
    \label{eq:split-spatial-eig}
\end{align}
Thus this mode is stable when $\cos\delta>1/2$ and unstable when $\cos\delta<1/2$.
Using Eq.~\eqref{eq:split-branch}, the zero occurs at
\begin{align}
    \gamma=3\mu .
\end{align}
For $\mu=1$, this predicts the observed split-1 to split-2 change near $\gamma=3$ in Fig.~\ref{fig:order-pars}.
At the representative split point $\gamma=2.5$, Eq.~\eqref{eq:split-spatial-eig} gives $\lambda_{\rm sp}=-5/18$, matching the largest nonneutral eigenvalue above.

The glass order parameters detect separation of the sign groups.
When all sign groups occupy the same point, $u_\pm=v_\pm=0$ and the state is ordinary sync.
When the sign groups separate, $u_\pm$ and $v_\pm$ become nonzero even if the rainbow order remains large.
Split 1 keeps the $+$ and $-$ sectors nearly equivalent; split 2 breaks that equivalence; splay is the limiting case in which one transformed coordinate remains clustered and the other spreads.

We do not have a closed-form expression for the later split and splay branches.
For prepared balanced signs, however, Eqs.~\eqref{eq:four-group-x}--\eqref{eq:four-group-theta} reduce them to a small algebraic fixed-point problem.
For the numerical split-2 and splay states in Fig.~\ref{fig:static-states}, the minimum within-group resultant in both $\xi$ and $\eta$ is one to machine precision, and the maximum residuals of the four-group equations are $4.9\times10^{-15}$ and $4.3\times10^{-15}$, respectively.
The phase-wave panel is different: the locked transformed coordinate is group-collapsed, but the spread coordinate is not.
It is nevertheless tied to a second exact branch.
Lock the $+$ sector and define
$z^-_{ab}=\langle e^{\ii\eta}\rangle_{(p,q)=(a,b)}$.
The antiphase phase wave
\begin{align}
    \xi_i&=\Xi,\qquad
    z^-_{++}=e^{\ii\psi},\qquad
    z^-_{--}=-e^{\ii\psi},\notag\\
    z^-_{+-}&=z^-_{-+}=0
    \label{eq:static-antiphase-pw}
\end{align}
is an exact fixed point for all $\mu$ and $\gamma$.
It predicts $r=1$, $s=0$, $u_+=v_+=0$, and $u_-=v_-=1/2$; the $-$-locked branch follows by interchanging $+$ and $-$.
The finite-$N$ linearization has global and spread-sector neutral modes.
After removing them, the representative point $(\mu,\gamma)=(-0.75,4)$ has largest real part $-2.16\times10^{-2}$.
A finite-$N$ grid check over $\gamma=0,0.1,1,4,8$ finds stable spectra at $\mu=-0.75,-0.25$ and unstable spectra at $\mu=-1.2,0.2$.
Thus, for $\gamma\ge0$, the branch explains the order-parameter plateau in Fig.~\ref{fig:static-states} and supports the observed static phase-wave window $-1<\mu<0$.
For iid signs, finite-size imbalance moves the branch locations and adds the $O(N^{-1/2})$ glass background seen in Sec.~\ref{sec:iid-robustness}.

\subsection{Async and Active Async}

At $\gamma=0$, Eqs.~\eqref{eq:xi}--\eqref{eq:eta} reduce to the ordinary ring model.
The async state is the uniform state in both transformed coordinates, $r=s=0$.
In that limit, the rainbow stability is controlled by the self-coupling $(1+\mu)/2$, so the ordinary async regime lies on the negative side of $\mu=-1$ in the identical-frequency model.

The continuum linearization is still simple because only the first Fourier modes enter the velocity field.
Treat the four sign classes as four populations with densities uniform in $(\xi,\eta)$ at async.
For one transformed coordinate, let
$z_a=\langle e^{\ii\zeta}\rangle_a$ be the first moment of sign group $a$, where $\zeta$ is either $\xi$ or $\eta$.
Let $a_0=(1+\mu)/2$, $c=\gamma/4$, and define the sector moments
\begin{align}
    \mathcal{R}=\langle z\rangle,\qquad
    \mathcal{U}=\langle pz\rangle,\qquad
    \mathcal{V}=\langle qz\rangle .
\end{align}
The corresponding first-harmonic velocity field has complex amplitude
\begin{align}
    H_a=a_0\mathcal{R}+c(p_a\mathcal{V}+q_a\mathcal{U}).
\end{align}
Since $\dot\zeta=\operatorname{Im}(H_ae^{-\ii\zeta})$, the first moment obeys
\begin{align}
    \dot z_a=\frac{1}{2}\left[
        a_0\mathcal{R}+c(p_a\mathcal{V}+q_a\mathcal{U})
        \right]
    \label{eq:group-first-harmonic}
\end{align}
to linear order.
Terms involving the opposite transformed coordinate average to zero against the uniform density, so the $+$ and $-$ sectors decouple.
For the balanced case, the sign moments obey $\langle p\rangle=\langle q\rangle=0$ and $\langle pq\rangle=\chi$.
Taking the $1$, $p$, and $q$ moments of Eq.~\eqref{eq:group-first-harmonic} gives two identical blocks,
\begin{align}
    Z_\pm&=(\mathcal{R}_\pm,\mathcal{U}_\pm,\mathcal{V}_\pm)^T,\notag\\
    \dot Z_\pm =
    \begin{pmatrix}
        a_0/2 & 0 & 0\\
        0 & c\chi/2 & c/2\\
        0 & c/2 & c\chi/2
    \end{pmatrix}Z_\pm,
    \label{eq:async-linear}
\end{align}
where $\chi=N^{-1}p^Tq$ is the sign overlap.
The rainbow eigenvalue and the two glass eigenvalues are
\begin{align}
    \sigma_{\rm r} &= \frac{1+\mu}{4},\\
    \sigma_{\rm g}^{\pm} &=
    \frac{\gamma}{8}(\chi\pm1).
    \label{eq:async-eigs}
\end{align}
The van Hemmen perturbation is therefore singular at $\gamma=0$: sign-weighted first-harmonic perturbations are neutral in the ordinary ring model, because the dynamics cannot see the labels $p_i,q_i$, but one glass combination grows as soon as $\gamma>0$ when $|\chi|<1$.
For the prepared balanced signs used in the figures, $\chi=0$, so $\sigma_{\rm g}^+=\gamma/8$ and $\sigma_{\rm g}^-=-\gamma/8$.

\begin{figure}[tbp]
\centering
\includegraphics[width=0.90\columnwidth]{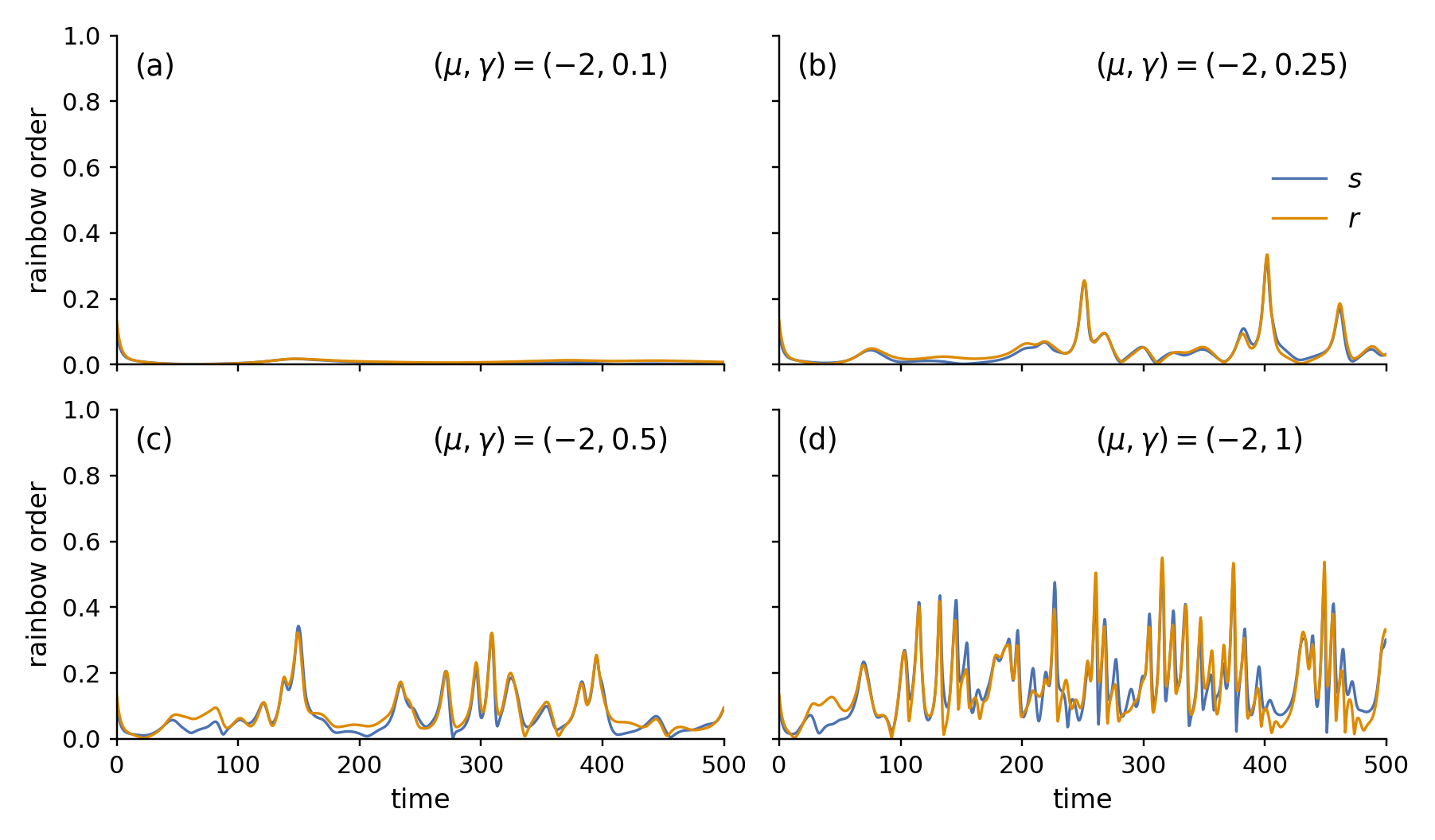}
\caption{Time series of the rainbow order parameters in the active async regime. The mean coupling is fixed at $\mu=-2$ and the disorder strength increases from panel (a) to panel (d), as labeled. The bursts grow in both amplitude and frequency as the van Hemmen part of the coupling is strengthened. Simulation details: $(dt,T,N)=(0.5,500,100)$ with seed $0$.}
\label{fig:bursty}
\end{figure}

The time series in Fig.~\ref{fig:bursty} are at $\mu=-2$, where $\sigma_{\rm r}=-1/4$ but the glass mode grows.
Thus weak van Hemmen disorder pulls the system off the static async manifold even when rainbow order is linearly damped.
The calculation detects the onset but not the nonlinear bursts, so we treat active async as a numerical state and leave its reduced dynamics open.
The checks in Appendix~\ref{app:validation} find nonzero tail mean speed across seeds, system sizes, and a halved time step, while $r_{\rm hi}$ remains small on average but bursts near $0.5$.

\subsection{Glassy Phase Wave}

\begin{figure}[tbp]
\centering
\includegraphics[width=0.90\columnwidth]{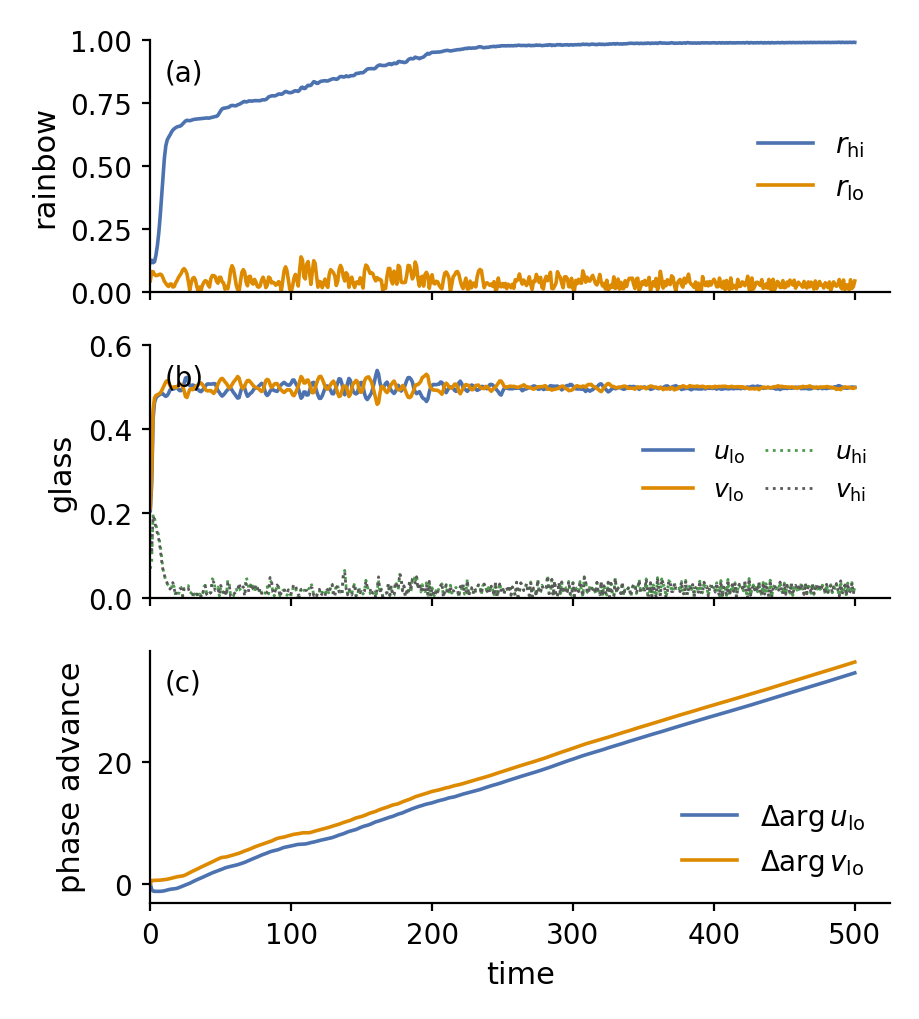}
\caption{Order parameters in the glassy phase wave. The magnitudes $u_{\rm lo}$ and $v_{\rm lo}$ settle rapidly, while $r_{\rm hi}$ relaxes on a longer time scale. The unwrapped phase advances $\Delta\arg u_{\rm lo}$ and $\Delta\arg v_{\rm lo}$ show the rotation of the low-sector glass order. Simulation details: $(\mu,\gamma)=(-1,8)$ and $(dt,T,N)=(0.1,500,64)$.}
\label{fig:glassy}
\end{figure}

The glassy phase wave is easier to diagnose because a simple identity constrains where glass order can live.
For the prepared equal-population sign pattern, define the missing Walsh component
\begin{align}
    R_\pm=\frac{1}{N}\sum_{j=1}^N p_jq_j e^{\ii(x_j\pm\theta_j)}.
\end{align}
Let $w_+=r$ and $w_-=s$.
If
\begin{align}
    z^\pm_{ab}=\frac{4}{N}\sum_{j:p_j=a,\,q_j=b}
    e^{\ii(x_j\pm\theta_j)},\qquad a,b=\pm1,
\end{align}
are the four sign-sector first moments, then Walsh orthogonality gives the exact identity
\begin{align}
    w_\pm^2+u_\pm^2+v_\pm^2+\lvert R_\pm\rvert^2
    =\frac{1}{4}\sum_{a,b=\pm1}\lvert z^\pm_{ab}\rvert^2\le1.
    \label{eq:walsh-bound}
\end{align}
The inequality is saturated only when every sign sector is internally collapsed in the corresponding transformed coordinate.
For each transformed coordinate,
the four complex averages $w_\pm e^{\ii\Phi_\pm}$, $u_\pm e^{\ii\Psi_\pm}$,
$v_\pm e^{\ii\Theta_\pm}$, and $R_\pm$ are therefore not independent.
Exact locking of one transformed coordinate would force the corresponding glass order parameters to vanish.
The large glass order in Fig.~\ref{fig:glassy} therefore appears in the low-rainbow sector, not in the nearly locked high-rainbow sector.

This gives a useful limiting reduction.
Suppose the high-rainbow coordinate is exactly locked and, without loss of generality, take it to be the $+$ sector.
For balanced signs this forces $u_+=v_+=0$.
The low coordinate $\zeta_i=\eta_i$ then obeys
\begin{align}
    \dot\zeta_i
    &=\frac{1+\mu}{2}
      \operatorname{Im}(w e^{\ii\Phi}e^{-\ii\zeta_i})\notag\\
    &\quad{}
      +\frac{\gamma}{4}
      \operatorname{Im}\{(p_i v e^{\ii\Theta}
      +q_i u e^{\ii\Psi})e^{-\ii\zeta_i}\},
    \label{eq:gpw-low-sector}
\end{align}
where $w,u,v$ are the low-sector rainbow and glass order parameters.
The uniform low-sector state has first-harmonic growth rates
\begin{align}
    \sigma_{\rm r}^{\rm lo}=\frac{1+\mu}{4},\qquad
    \sigma_{\rm g}^{\rm lo,\pm}=\frac{\gamma}{8}(\chi\pm1).
    \label{eq:gpw-low-sector-eigs}
\end{align}
Thus at the plotted point $(\mu,\gamma)=(-1,8)$ with $\chi=0$, the low-rainbow mode is not linearly amplified while the glass mode grows at rate one.
This explains the rapid onset of $u_{\rm lo},v_{\rm lo}$ with small $r_{\rm lo}$.

The large-glass limit of Eq.~\eqref{eq:gpw-low-sector} is also explicit.
With sign-group first moments
$z_{pq}=\langle e^{\ii\zeta}\rangle_{(p,q)}$, the state
\begin{align}
    z_{++}=e^{\ii\psi},\qquad
    z_{--}=-e^{\ii\psi},\qquad
    z_{+-}=z_{-+}=0
    \label{eq:gpw-antiphase-glass}
\end{align}
has $w=0$, $u=v=e^{\ii\psi}/2$, and $R_{\rm lo}=0$.
It is a one-parameter family of fixed points in the locked-high-sector approximation.
The phase $\psi$ is therefore neutral at leading order.
The rotation in Fig.~\ref{fig:glassy} is a slow drift of this neutral glass phase, produced by the finite high-sector relaxation and the small residual terms absent from Eq.~\eqref{eq:gpw-low-sector}.

It has the order-parameter signature
\begin{align}
    r_{\rm hi}\ \text{large},\qquad
    r_{\rm lo}\simeq0,\qquad
    u_{\rm lo}\simeq v_{\rm lo}>0,
\end{align}
with the high-sector glass order parameters small.
Thus one transformed coordinate is phase-wave-like and nearly locked, while the spread coordinate has little ordinary rainbow order but remains organized by the quenched signs.
The macroscopic order is therefore glassy in the van Hemmen sense: it is sign-weighted, not just rainbow order.

The striking feature in Fig.~\ref{fig:glassy} is the separation of time scales.
The low-sector glass magnitudes rapidly plateau, while $r_{\rm hi}$ relaxes slowly and the glass phases rotate.
In the matched immobile Kuramoto-van Hemmen control with $\omega_i=0$, the symmetric coupling gives a gradient system, and the static checks behind Table~\ref{tab:state-comparison} find locked or glass fixed points.
The finite-width Kuramoto problem has the richer incoherent, partial-locking, antiphase, and mixed taxonomy of Ref.~\cite{kloumann2013phase}.
Here the same type of quenched disorder can support a rotating glass order parameter because the phases and positions continually reshape one another.

The basin is finite.
At $(\mu,\gamma)=(-1,8)$, three of five random initial conditions reached the high-rainbow branch by $T=500$; all five developed sign-weighted low-sector order.
The seed-$0$ high-rainbow branch persisted under smaller $dt$, longer integration, and $N=128$ checks; details are in Appendix~\ref{app:validation}.

\section{Discussion}

\begin{table}[tbp]
\caption{Matched $\omega_i=0$ Kuramoto--VH control versus the swarmalator model. For the Kuramoto column, $R=|\langle e^{\ii\theta}\rangle|$, $P=|\langle p e^{\ii\theta}\rangle|$, and $Q=|\langle q e^{\ii\theta}\rangle|$. A \texttt{--} entry means no matched counterpart is claimed here.}
\label{tab:state-comparison}
\footnotesize
\setlength{\tabcolsep}{3pt}
\begin{tabular}{@{}>{\raggedright\arraybackslash}p{0.30\columnwidth}
                >{\raggedright\arraybackslash}p{0.24\columnwidth}
                >{\raggedright\arraybackslash}p{0.34\columnwidth}@{}}
\hline\noalign{\smallskip}
Kuramoto-VH, $\omega_i=0$ & Swarmalator-VH & Diagnostic/status \\
\noalign{\smallskip}\hline\noalign{\smallskip}
Sync & Sync & locked; $R=1$ and $r=s=1$ \\
Split glass & Split 1 & same branch; $\cos\delta=\mu/(\gamma-\mu)$ \\
Antiphase glass & \texttt{--} & static; $R=0$, $P=Q=1/2$ \\
\texttt{--} & Split 2 & transformed sectors separate \\
\texttt{--} & Splay & sign groups spread one sector \\
\texttt{--} & Phase wave & one sector locked, one spread; $m=0$ \\
\texttt{--} & Active async & bursty small $r,s$, $m>0$ \\
\texttt{--} & Glassy phase wave & low-sector glass, $m>0$ \\
\noalign{\smallskip}\hline
\end{tabular}
\end{table}

We introduced a one-dimensional swarmalator model with van Hemmen pair interactions.
In transformed coordinates the model closes through six complex order parameters, but it is not a copy of the Kuramoto-van Hemmen problem.
The matched zero-frequency Kuramoto control has one ordinary and two glass order parameters, and its balanced fixed points include sync, the same first split-glass branch, and static antiphase-glass configurations.
The familiar incoherent, partially locked, antiphase, and mixed labels in Ref.~\cite{kloumann2013phase} belong to the finite-frequency-width problem rather than to this identical-oscillator control.
Here phase changes reshape future spatial interactions, and spatial rearrangements reshape future phase interactions, doubling the natural order-parameter set and supporting active states even with identical natural frequencies.
Table~\ref{tab:state-comparison} summarizes the closest analogues.

The sync boundary $\gamma_c(\chi)=2\mu/(1+\chi)$, with large-$N$ limit $2\mu$, is the sharpest bifurcation result.
It follows from the low-rank structure of the van Hemmen coupling matrix and matches the numerical transition from sync to split states.
The first split branch adds a second explicit prediction: it has $r=s=(1+\cos\delta)/2$, $u_\pm=v_\pm=|\sin\delta|/2$, and a pure-spatial eigenvalue that changes sign at $\gamma=3\mu$.
The antiphase phase-wave branch gives another exact prediction: one transformed coordinate locks, the other has $r_{\rm lo}=0$ but $u_{\rm lo}=v_{\rm lo}=1/2$, and transverse stability checks support the static window $-1<\mu<0$ for $\gamma\ge0$.
Pair disorder also creates static glass order, organizes the phase-wave sector by quenched signs, and supports active regimes whose macroscopic order parameters do not settle to constants.
For prepared balanced signs, split and splay are numerically exact four-group fixed points.
For fully random signs their stability remains open.
So do the nonlinear burst mechanism in active async and a self-consistency theory for the glassy phase wave.
Fully random pair couplings, distributed natural frequencies, and noisy interactions are natural next steps.

\appendix

\section{Numerical Validation}
\label{app:validation}

This appendix records the checks behind the state classifications.
All checks use the same equations and fourth-order Runge--Kutta scheme as the figures; tail averages use the final fifth of each sampled trajectory.
The transformed-reduction script compares Eqs.~\eqref{eq:xi}--\eqref{eq:eta} against the original $(x,\theta)$ equations on random prepared and iid sign states.

For static states, the validation script substitutes the stored representatives into the full $2N$-dimensional equations and, when applicable, into the four-group quotient.
The sync and first split branches have maximum full-system residuals below $10^{-15}$ and $4\times10^{-16}$, respectively.
The split stability script verifies the exact spatial eigenmode in Eq.~\eqref{eq:split-spatial-eig} and its sign change at $\gamma=3\mu$.
The split-2 and splay states have maximum full-system residuals below $5\times10^{-15}$, within-group resultants equal to one to machine precision, and four-group residuals $4.9\times10^{-15}$ and $4.3\times10^{-15}$.
The phase-wave representative has maximum speed $1.1\times10^{-10}$; only the locked transformed coordinate collapses by sign group.
The exact antiphase phase-wave check constructs Eq.~\eqref{eq:static-antiphase-pw}, verifies $r_{\rm hi}=1$, $r_{\rm lo}=0$, and $u_{\rm lo}=v_{\rm lo}=1/2$, and finds largest nonzero real part $-2.16\times10^{-2}$ at $(\mu,\gamma)=(-0.75,4)$.
The phase-wave stability-grid script checks the same branch over $\gamma=0,0.1,1,4,8$: sampled points inside $-1<\mu<0$ are stable, while sampled points at $\mu=-1.2$ and $\mu=0.2$ are unstable.

The iid-sign checks in Sec.~\ref{sec:iid-robustness} were produced by \texttt{audit\_iid\_signs.py}.
The script computes exact finite-$N$ sync thresholds by bisection on the largest nonneutral eigenvalue of Eq.~\eqref{eq:B} and performs the random-start static sweep summarized there.

The immobile comparison in Table~\ref{tab:state-comparison} was checked with the zero-frequency Kuramoto script, using the matched control
$\dot\theta_i=N^{-1}\sum_jK_{ij}\sin(\theta_j-\theta_i)$ and the same prepared balanced signs.
For $\mu>0$, sync loses stability at $\gamma=2\mu$ and the first glass branch has
$R=(1+\cos\delta)/2$, $P=Q=|\sin\delta|/2$, and $\cos\delta=\mu/(\gamma-\mu)$.
The same script verifies an exact static antiphase-glass configuration with $R=0$ and $P=Q=1/2$.

For the active async representative, five random initial conditions at $(\mu,\gamma)=(-2,1)$ with $N=100$, $dt=0.5$, and $T=500$ all remained active: tail mean speeds ranged from $0.126$ to $0.197$, tail mean $r_{\rm hi}$ from $0.179$ to $0.226$, tail maxima of $r_{\rm hi}$ from $0.49$ to $0.56$, and the fraction of sampled times with $r_{\rm hi}>0.2$ from $0.34$ to $0.50$.
A resolution check at seed $0$ gave the same classification for $N=64,100,128$ and for halving the step from $dt=0.5$ to $0.25$: final-tail mean speeds lay between $0.126$ and $0.158$, and the tail maxima of $r_{\rm hi}$ lay between $0.50$ and $0.55$.

For the glassy phase wave at $(\mu,\gamma)=(-1,8)$, the seed ensemble shows multistability: three of five random initial conditions reached the high-rainbow branch by $T=500$, with tail-averaged $r_{\rm hi}>0.95$ and $r_{\rm lo}<0.12$, while the other two stayed in intermediate moving states with $r_{\rm hi}\simeq0.39$ and $r_{\rm lo}\simeq0.27$.
All five runs developed sign-weighted low-sector order: the tail average of $(u_{\rm lo}+v_{\rm lo})/2$ lay between $0.38$ and $0.50$.
A prepared-sign sweep at $\mu=-1$, $N=64$, and seed $0$ found the high-rainbow branch for $\gamma\ge2.5$, with $0.960<r_{\rm hi}<0.987$ and $0.038<r_{\rm lo}<0.083$.
For the seed-$0$ high-rainbow branch, the same signature persisted when $dt$ was halved, when the integration time was extended to $T=1000$, and when the system size was increased to $N=128$.
Across those checks, the tail means obeyed $0.90<r_{\rm hi}<0.995$, $0.023<r_{\rm lo}<0.036$, and $(u_{\rm lo}+v_{\rm lo})/2\simeq0.499$.
The Fig.~\ref{fig:glassy} generation script also checks that the low-sector glass phases advance by more than one full turn in the plotted run.

\section*{Data Availability}

The reproducibility archive contains the code, data, figure scripts, manuscript source, and validation suite needed to reproduce the numerical results.
It includes the fast build checks and the slower unsteady and iid-sign audits reported in Appendix~\ref{app:validation}.
A frozen public copy of the archive will be deposited before publication.


\end{document}